\documentclass[sn-standardnature]{sn-jnl}

\usepackage{float}
\usepackage{hyperref}
\PassOptionsToPackage{numbers, sort&compress}{natbib}
\usepackage{siunitx}
\usepackage{afterpage}
\usepackage{multirow}
\begin{document}

\title[Article Title]{Generating counterfactual explanations of tumor spatial proteomes to discover effective strategies for enhancing immune infiltration}

\author[1]{\fnm{Zitong Jerry} \sur{Wang}}
\author[2]{\fnm{Alexander} \sur{M. Xu}}
\author[1]{\fnm{Aman} \sur{Bhargava}}
\author*[1]{\fnm{Matt} \sur{W. Thomson}}\email{mthomson@caltech.edu}

\affil[1]{\orgdiv{Division of Biology and Biological Engineering}, \orgname{California Institute of Technology}, \orgaddress{\street{1200 E California Blvd}, \city{Pasadena}, \postcode{91125}, \state{CA}, \country{USA}}}
\affil[2]{\orgdiv{Samuel Oschin Comprehensive Cancer Institute}, \orgname{Cedars-Sinai Medical Center}, \orgaddress{\street{8700 Beverly Blvd}, \city{Los Angeles}, \postcode{90048}, \state{CA}, \country{USA}}}

\abstract{The tumor microenvironment (TME) significantly impacts cancer prognosis due to its immune composition. While therapies for altering the immune composition, including immunotherapies, have shown exciting results for treating hematological cancers, they are less effective for immunologically-cold, solid tumors. Spatial omics technologies capture the spatial organization of the TME with unprecedented molecular detail, revealing the relationship between immune cell localization and molecular signals. Here, we formulate T-cell infiltration prediction as a self-supervised machine learning problem and develop a counterfactual optimization strategy that leverages large scale spatial omics profiles of patient tumors to design tumor perturbations predicted to boost T-cell infiltration. A convolutional neural network predicts T-cell distribution based on signaling molecules in the TME provided by imaging mass cytometry. Gradient-based counterfactual generation, then, computes perturbations predicted to boost T-cell abundance. We apply our framework to melanoma, colorectal cancer (CRC) liver metastases, and breast tumor data, discovering combinatorial perturbations predicted to support T-cell infiltration across tens to hundreds of patients. This work presents a paradigm for counterfactual-based prediction and design of cancer therapeutics using spatial omics data.}


\maketitle

\section*{Introduction}

The immune composition of the tumor microenvironment (TME) plays a crucial role in determining patient prognosis and response to cancer immunotherapies \cite{fridman2017immune, binnewies2018understanding, bruni2020immune}. Immunotherapies that alter the immune composition using transplanted or engineered immune cells (chimeric antigen receptor T cell therapy) or remove immunosuppressive signaling (checkpoint inhibitors) have shown exciting results in relapsed and refractory tumors in hematological cancers and some solid tumors. However, effective therapeutic strategies for most solid tumors remain limited \cite{hegde2020top,choe2020engineering,pitt2016targeting}. The TME is a complex mixture of immune cells, including T cells, B cells, natural killer cells, and macrophages, as well as stromal cells and tumor cells \cite{fridman2017immune}. The interactions between these cells can either promote or suppress tumor growth and progression, and ultimately impact patient outcomes. For example, high levels of tumor-infiltrating lymphocytes (TILs) in the TME are associated with improved prognosis and response to immunotherapy across multiple cancer types \cite{haslam2019estimation,lee2019multiomics}. Conversely, an immunosuppressive TME, characterized by low levels of TILs and high levels of tumor-associated macrophages (TAMs), is associated with poor prognosis and reduced response to immunotherapy \cite{pittet2022clinical}. Durable, long-term clinical response of T-cell-based immunotherapies are often constrained by a lack of T-cell infiltration into the tumor, as seen in classically ``cold" tumors such as triple-negative breast cancer (TNBC) or prostate cancer, which have seen little benefit from immunotherapy \cite{bonaventura2019cold, savas2016clinical, tsaur2021immunotherapy}.

Spatial omics technologies capture the spatial organization of cellular activity in intact human tumors with unprecedented molecular detail, revealing the relationship between localization of different cell types and tens to thousands of molecular signals \cite{moffitt2022emerging}, paving the way for the design of novel strategies to manipulate the tumor immune microenvironment. T-cell infiltration is modulated by a rich array of signals within the tumor microenvironment (TME) such as chemokines, adhesion molecules, tumor antigens, immune checkpoints, and their cognate receptors \cite{lanitis2017mechanisms}. Recent advances in in situ molecular profiling techniques, including spatial transcriptomic \cite{rodriques2019slide,eng2019transcriptome} and protemic \cite{giesen2014highly,goltsev2018deep} techniques, simultaneously captures the spatial relationship of tens to thousands of molecular signals and T cell localization in intact human tumors with micron-scale resolution. Imaging mass cytometry (IMC) combines immunohistochemistry using metal-labeled antibodies with laser ablation and detection using mass cytometry by time-of-flight to enable simultaneous detection of up to 40 antigens and transcripts in intact tissue \cite{giesen2014highly}.

Recent work on computational methods as applied to multiplexed tumor images have primarily focused on predicting patient-level phenotypes such as survival, by identifying spatial motifs from tumor microenvironments \cite{bhate2022tissue, wu2022graph, schurch2020coordinated}. These methods have generated valuable insights into how the structure of TMEs influences patient prognosis and treatment response, but they fall short of generating concrete, testable hypothesis for therapeutic interventions that may improve patient outcome. Given the importance of immune cell localization such as CD8+ T cells in determining patient outcomes, we need computational tools that can predict immune cell localization from environmental signals and systematically generate specific, feasible tumor perturbations that are predicted to alter the TME in a way that improves patient outcome.

In this work, we formulated T-cell infiltration prediction as a self-supervised machine learning problem, and critically, we combine this prediction task with the approach of counterfactual explanations to design perturbations predicted to increase T-cell infiltration.  
Specifically, we first train a convolutional neural network to predict T-cell level based upon micron-resolution maps of the signaling molecules in the TME provided by IMC. We then apply gradient-based counterfactual generation strategy to the infiltration neural network to compute perturbations to the signaling molecules that the network predicts can increase T-cell abundance.
We apply our analysis framework to spatial proteomic profiles of tumors from melanoma \cite{hoch2022multiplexed}, CRC liver metastases \cite{wang2023extracellular} and breast cancer patients \cite{danenberg2022breast}, discovering tumor perturbations that are predicted to support T cell infiltration across tens to hundreds of patients. In melanoma patients, our framework predicts combinatorial perturbation to the tumoral level of CXCL9, CXCL10, CCL22 and CCL18 can convert immune-excluded tumors to be immune-inflamed in a cohort of 69 patients. For CRC liver metastasis, our framework discovered a therapeutic strategy consisting of CXCR4, PD1, PDL1 and CYR61 that is predicted to consistently improve T-cell infiltration across a cohort of 30 patients. Our work provides a paradigm for counterfactual-based prediction and design of cancer therapeutics based on classification of immune system activity in spatial proteomics and transcriptomic data.

\section*{Results}

\subsection*{A counterfactual optimization framework for in situ molecular profiles enables the discovery of tumor perturbations predicted to drive T cell infiltration}

\begin{figure}[t!]
  \centering
  \includegraphics[width=\textwidth]{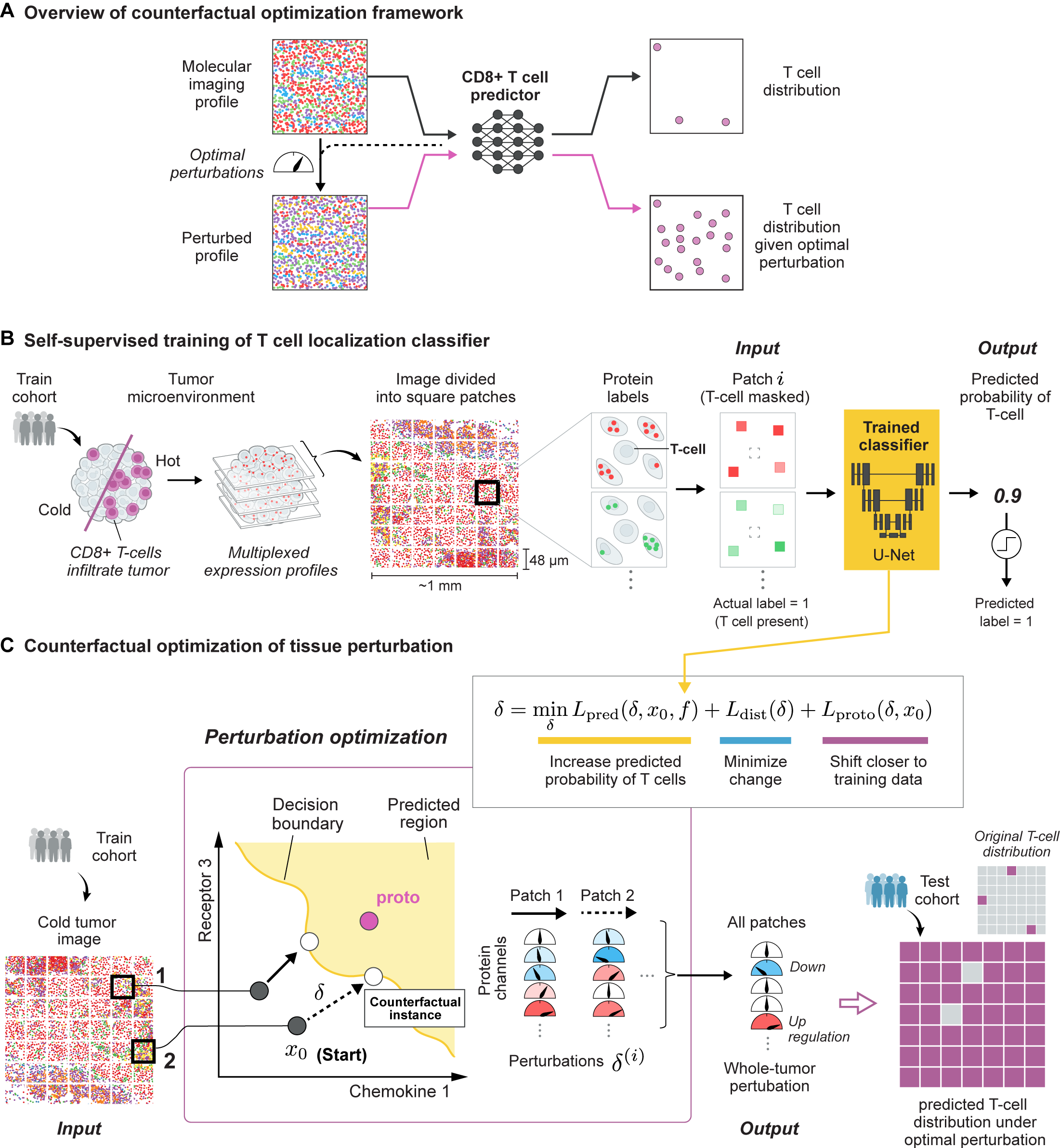}
  \caption{(A) Overview of counterfactual optimization framework for discovering therapeutic strategies predicted to support CD8+ T cell infiltration in immune-cold tumors. Framework consists of first (B) training a classifier (neural network) to predict the presence of CD8+ T cells from multiplexed tissue images where CD8+ T cells are masked. (C) The trained classifier is then used to compute an optimal perturbation vector $\delta$ per patch, which represents a strategy for altering the level of different proteins in a way that increases the probability of T cell presence as predicted by the classifier. The positive/negative values (shaded red/blue) of $\delta$ correspond to increases/decreases in mean relative channel intensity. Finally, taking the median across all such perturbations enables us to obtain a tumor-wide perturbation vector that represents a more feasible and potentially effective therapeutic strategy.}
  \label{framework}
\end{figure}

The general logic of our framework is to train, in a self-supervised manner, a classifier to predict the presence of CD8+ T cells from tissue images with CD8+ T cells masked (\autoref{framework}A,B). Then we apply counterfactual reasoning to the data by performing gradient descent on the input image, allowing us to discover perturbations to the tumor image that increases the classifier's predicted likelihood of CD8+ T cells being present (\autoref{framework}A,C). The altered image represents a perturbation of the TME predicted to improve T cell infiltration. 


We leverage IMC profiles of human tumors to train a classifier to predict the spatial distribution of CD8+ T cell in a self-supervised manner. Consider a set of IMC image patches $\{I^{(i)}\}$ where $I^{(i)} \in \mathbb{R}^{l \times w \times c}$, with $l$ and $w$ denotes the pixel length and width of the image and $c$ denotes the number of molecule channels in the images (\autoref{framework}B). Each image shows the level of $c$ proteins across all cells within a small patch of tissue. From patch $I^{(i)}$, we obtain a binary label $s^{(i)}$ indicating the presence and absence of CD8+ T cells in the patch and a masked copy $x^{(i)}$ with all signals originating from CD8+ T cells removed (see \nameref{sec:methods}). The task for our model $f$ is to classify whether T cells are present ($s^{(i)}=1$) or absent ($s^{(i)}=0)$ in image $I^{(i)}$ using only its masked copy $x^{(i)}$. Specifically, $f(x^{(i)}) \in [0,1]$ is the predicted probability of T cells, and then we apply a classification threshold $p$ to convert this probability to a predicted label $\hat{s}^{(i)} \in \{0, 1\}$. Since we obtain the image label $s^{(i)}$ from the image $I^{(i)}$ itself by unsupervised clustering of individual cells, our overall task is inherently self-supervised.

Given a set of image patches, we train a model $f$ to minimize the following T cell prediction loss, also known as the binary cross entropy (BCE) loss,
\begin{equation}
	L = - \frac{1}{N} \sum_{i=1}^{N} \left[ s^{(i)} \log\left(\hat{s}^{(i)}\right) + \left(1 - s^{(i)}\right) \log\left(1 - \hat{s}^{(i)}\right) \right],
\end{equation}
where 
\begin{equation}
\hat{s}^{(i)} = 
\begin{cases} 
1 & \text{if } f(x^{(i)}) \geq p \\
0 & \text{if } f(x^{(i)}) < p
\end{cases}
\end{equation}
and $p$ is the classification threshold. We select $p$ by minimizing the following root mean squared error (RMSE) on a separate set of tissue sections $\Omega$,

\begin{equation}
\text{RMSE}^2 = \frac{1}{\lvert\Omega\rvert}\sum_{j \in \Omega}\left\| \frac{1}{N_j} \sum_{i=1}^{N_j} \hat{s}^{(i)} - s^{(i)} \right\|^2.
\label{eq:RMSE}
\end{equation}

The RMSE is a measure of the discrepancies between the observed and predicted proportions of T cell patches in a tissue section averaged across a set of tissues, which we take to be the validation set.

We analyzed the performance of a series of convolutional neural network (CNN) architectures as our classifier, including CNNs and vision transformers, and saw similar performance across the board (\autoref{tab:modelcomparison}). We settled on a U-Net architecture because of ease of extension of the model to multichannel data sets. Our U-Net classifier consists of a standard U-Net architecture \cite{buda2019association} and a fully-connected layer with softmax activation (see \nameref{sec:methods}). To increase the number of samples available for training, we take advantage of the spatial heterogeneity of TMEs and divide each tissue image into \SI{50}{\micro\meter} $\times$ \SI{50}{\micro\meter} patches upon which the classifier is trained to predict T cell presence (see \nameref{sec:methods} for training details).

Using our trained classifier and IMC profiles of cold tumors, we employ a counterfactual optimization method to predict tumor perturbations that enhance CD8+ T cell infiltration (\autoref{framework}C). For each image patch $x_0^{(i)}$ lacking CD8+ T cell, our optimization algorithm searches for a perturbation $\delta^{(i)}$ such that our classifier $f$ predicts the perturbed patch $x_p^{(i)} = x_0^{(i)} + \delta^{(i)}$ as having T cells, hence referred to as a counterfactual. In general, $\delta^{(i)}$ is a 3D tensor that describes perturbation made to each pixel of the patch, which we compute by solving the following optimization problem adopted from \cite{looveren2021interpretable},
\begin{equation}
	\delta^{(i)} = \min_{\delta} L_\mathrm{pred}(x_0^{(i)}, \delta) + L_\mathrm{dist}(\delta) + L_\mathrm{proto}(x_0^{(i)}, \delta), 
	\label{eq:objectivefun} 
\end{equation}
	such that 
\begin{equation}
\begin{split}
  L_\mathrm{pred}(x_0^{(i)}, \delta) &= c\max(-f(x_0^{(i)} + \delta),\, -p), \\
  L_\mathrm{dist}(\delta) &= \beta \lVert \delta \rVert_{1} + \lVert \delta \rVert_{2}^2, \\
  L_\mathrm{proto}(x_0^{(i)}, \delta) &= \theta  \lVert x_0^{(i)} + \delta -\mathrm{proto}^{(i)} \rVert_{2}^2 
\end{split}
\label{eq:objective_detail} 
\end{equation}
where $p$ is the classification threshold.

The three loss terms in \autoref{eq:objectivefun} each correspond to a desirable property of the perturbation we aim to discover. The $L_\mathrm{pred}$ term encourages the perturbation to increase the classifier's predicted probability of T cells to be larger than $p$, meaning the network will predict the perturbed tissue patch as having T cells when it previously did not contain T cells. Next, the elastic net regularization term, $L_\mathrm{dist}$, minimizes the distance between the original patch $x_0^{(i)}$ and the perturbed patch $x_\mathrm{p}^{(i)} = x_0^{(i)} + \delta$ to generate sparse perturbations, favoring strategies that do not require making as many changes to the tumor. Lastly, $\mathrm{proto}^{(i)}$ in $L_\mathrm{proto}$ refers to, among all patches in the training set that is classified as having T cells, the nearest neighbour of $x_0^{(i)}$ (see \nameref{sec:methods}). Thus this last term explicitly guides the perturbed image $x_\mathrm{p}^{(i)}$ to falls in the distribution of training images predicted to contain T cells, leading to perturbed patches that appear similar to what has been observed in TMEs infiltrated by T cells.

Since drug treatments cannot act at the spatial resolution of individual micron-scale pixels, we also constrain the spatial resolution of our perturbations by defining $\delta^{(i)}$ in the following way,
\begin{equation}
	\delta^{(i)} = \gamma^{(i)} \odot_3 x^{(i)}_0,
	\label{eq:deltaconstraint}
\end{equation}
where $\gamma^{(i)} \geq -1$ a $c$-dimensional vector with $c$ being the number of protein channels. In practice, we directly optimize for $\gamma^{(i)}$, where $\gamma^{(i)}_j$ can be interpreted as the relative change to the mean intensity of the $j$-th channel. This constraint ensures that the perturbation we find changes the level of any molecule by the same multiplicative factor across all cells. However, given our classifier does have fine spatial resolution, we can search for targeted therapies such as perturbing only a specific cell type or restricting the perturbation to specific tissue locations by changing \autoref{eq:deltaconstraint}.

Taken together, our algorithm obtains an altered image predicted to contain T cells from an original image which lacks T cells, by minimally perturbing the original image in the direction of the nearest training patch containing T cells until the classifier predicts the perturbed image to contain T cells. Since our strategy may find different perturbations for different tumor patches, we reduce the set of patch-wise perturbations $\{\delta^{(i)}\}_i$ to a whole-tumor perturbation by taking the median across the entire set.

\subsection*{Convolutional neural networks trained on IMC images accurately predict T cell distribution from distribution of other cell types}

\begin{figure}[H]
  \centering
  \includegraphics[width=\textwidth]{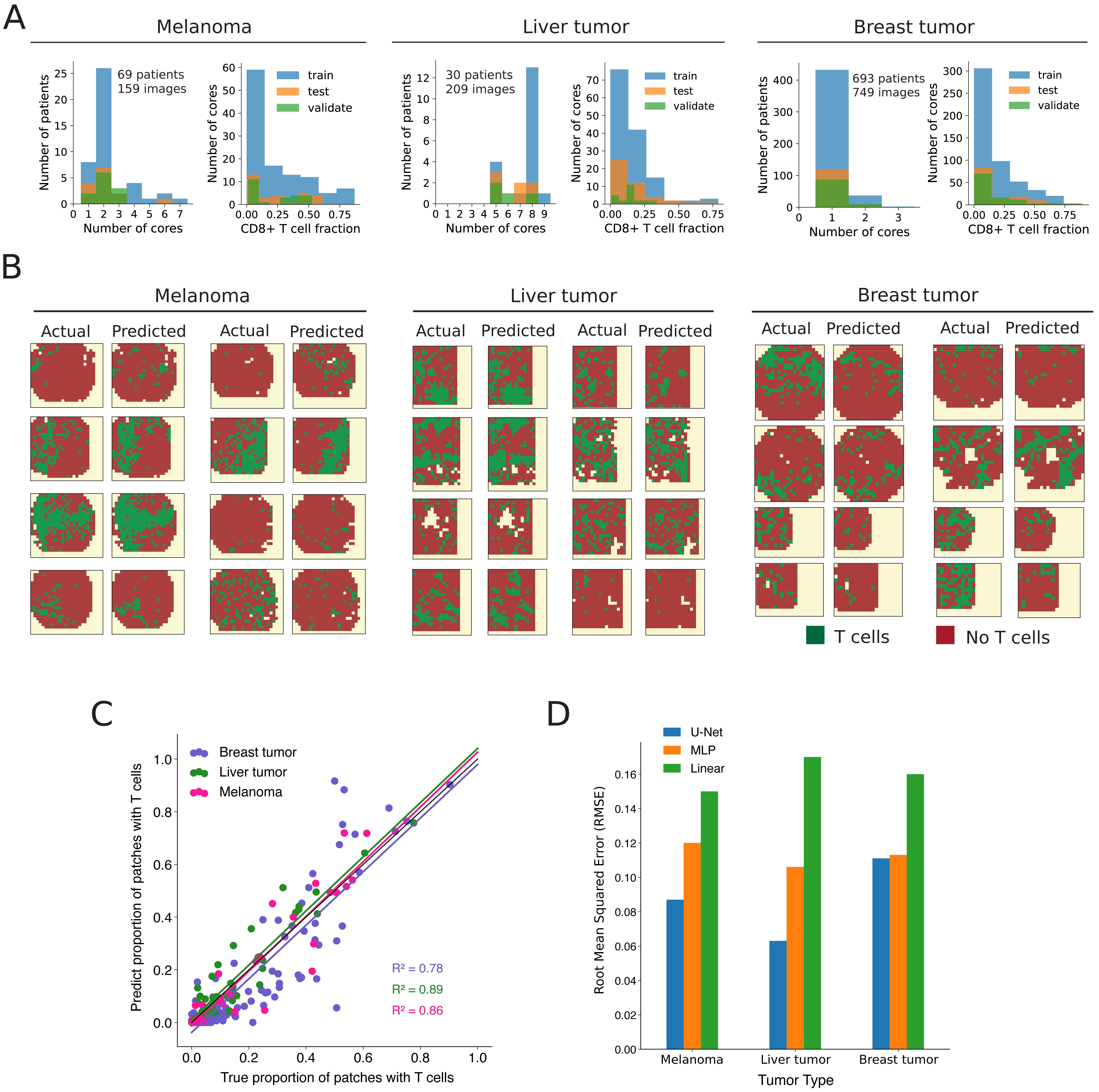}
  \caption{\textbf{U-Net classifier trained on IMC images of melanoma, metastatic liver , and breast tumor accurately predicts T cell distribution.} 
  (A) histograms showing the distribution of tumor cores per patient and CD8+ T cell fractions per core across all three tumor datasets. 
  (B) Predicted and actual T cell distribution of tissue sections from test patients in melanoma, liver tumor, and breast tumor dataset. 
  (C) Predicted and true proportion of patches with T cells within a tissue section, each dot corresponds to a tissue section, points along the diagonal black line indicates perfect prediction. Paired t-test on predicted and true proportion gives p-value of 0.87 (melanoma), 0.81(liver tumor), 0.90 (breast tumor), indicating the predicted and true proportions are not distinct enough to reject the null that they are represent the same population
  (D) The RMSE (\autoref{eq:RMSE}) across all tissue sections from test patients, prediction from three different types of models.}
  \label{fig:classifier_performance}
\end{figure}

We applied our optimization framework to three publicly available IMC datasets of tumors from metastatic melanoma \cite{hoch2022multiplexed}, liver metastasis generated by primary colorectal cancer, \cite{wang2023extracellular}, and breast cancer patients \cite{danenberg2022breast}. The melanoma dataset was obtained by IMC imaging of 159 tumor cores from 69 patients with stage III or IV metastatic melanoma with samples collected from sites including skin and lymph-node. Each tissue was approximately \qty{1}{\milli\meter} in diameter, imaged at \qty{1}{\micro\meter} resolution across 41 channels, consisting of protein markers for tumor cells, immune cells, stromal cells and $11$ different chemokines \cite{hoch2022multiplexed}. The liver metastasis dataset consists of 209 tumor cores taken from 30 colorectal cancer patients with liver metastasis \cite{wang2023extracellular} spanning patients with and without a diagnosis of non-alcoholic fatty liver disease. The breast cancer dataset was obtained by IMC imaging of $749$ breast tumor cores from 693 patients from the METABRIC study with different molecular sub-types of breast cancer \cite{danenberg2022breast} . Most tissues (93\%) were \qty{0.6}{\milli\meter} in diameter, image at 1um resolution across 37 channels, consisting of markers for lymphoid, myeloid and stromal cells.




For each of the three tumor datasets, we trained a separate U-net classifier that effectively predicts CD8+ T cell infiltration level in unseen tumor sections. \autoref{fig:classifier_performance}B shows example tissue sections colored by the predictions made by the U-net classifiers, with green denoting a patch was predicted to have T cells. For each tissue section of a cancer type, the predictions were obtained by applying the corresponding classifier to each image patch. From visual inspection, our model consistently captures the general distribution of T cells. Comparing the true proportion of patches in a tissue section with T cells against our predicted proportion also shows strong agreement (\autoref{fig:classifier_performance}C). The true proportion of patches with T cells is calculated by dividing the number of patches within a tissue section that contains CD8+ T cells by the total number of patches within that section, hence it is restricted to be between 0 and 1. We quantify the performance of our U-Nets on the test data set using the RMSE (\autoref{eq:RMSE}) which represents how far off, on average, our predicted proportion is from the true proportion. Our model performs exceptionally well on liver tumor and melanoma, achieving a RMSE of only $6\%$ and $8\%$ respectively and a relatively lower performance of $11\%$ on breast tumor(\autoref{fig:classifier_performance}D). Taken together, these results suggest that our models accurately predict the T cell infiltration status of melanoma, liver and breast tumors.

In order to gain insight into the relative importance of non-linearity and spatial information in the performance of the U-Net on the T cell clasification task, we compared the performance of the U-Net to two other models (A) a logistic regression model (B) a multi-layer perceptron (MLP). The logistic regression model is essentially a linear model with a threshold, and the MLP is a non-linear model without explicit spatial structures. \autoref{fig:classifier_performance}D shows that across all three cancer datasets, the MLP classifier consistently outperforms the logistic regression model, reducing RMSE by $20-40\%$ to suggest that there are significantly nonlinear interactions between different molecular features when it comes to predicting the presence of T cells. The importance of spatial features on the T cell prediction task, however, is less consistent across cancer types. \autoref{fig:classifier_performance}D shows that for the prediction task in breast tumor, the U-Net model offers negligible boost in performance relative to the MLP model ($<2\%$ RMSE reduction), whereas for liver tumor, the U-Net model achieved a RMSE $50\%$ lower  compared to the MLP model. This preliminary result suggests that the spatial organization of signals may have a stronger influence on CD8+ T cell localization in liver tumor compared to breast tumor, or that the functional role of signals profiled in the liver data sets are more spatially-dependent.

\subsection*{Optimization framework predicts combinatorial chemokine treatment can improve T cell infiltration in melanoma patients}

\begin{figure}[t!]
  \centering
  \includegraphics[width=\textwidth]{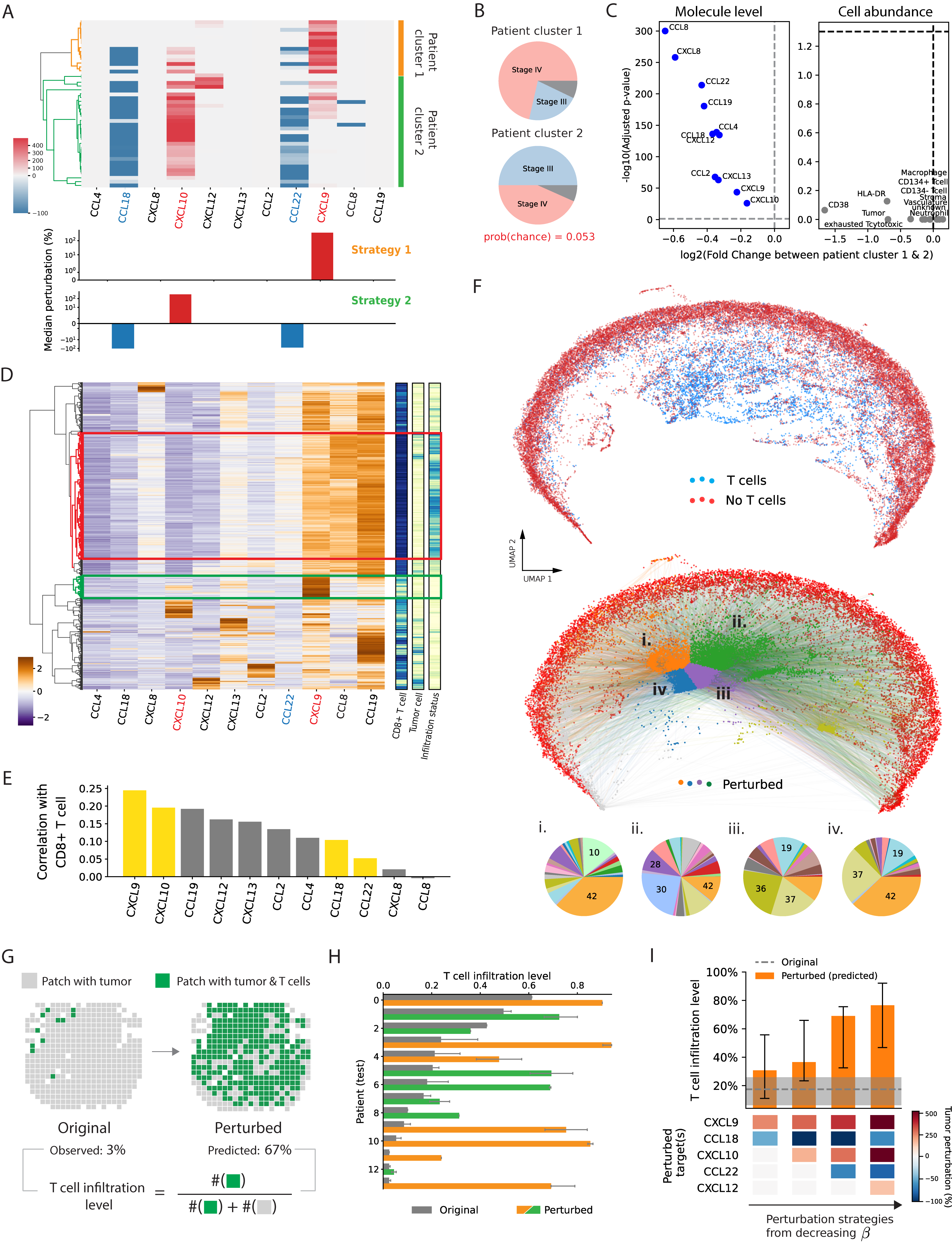}
  \caption{\textbf{Tumor perturbation strategy optimized on groups of melanoma patients} 
  (A) Chemokine perturbations computed for each patient (row) from training cohort; Bar graph shows the median perturbation across all patients, expressed in terms of the relative change in intensity.
  (B) Distribution of cancer stages among patients within two clusters, gray indicates unknown cancer stage. 
  (C) Volcano plot comparing chemokine level and cell type abundance from patient cluster 1 and 2, computed using median values and Wilcoxon rank sum test. Gray indicates non-statical significance, blue indicates highly enriched in patient cluster 2, red indicates highly enriched in patient cluster 1. 
  }
  \label{fig:melanoma}
\end{figure}

\addtocounter{figure}{-1}
\begin{figure}[t!]
    \centering
        \caption[]{\textbf{(continued)} (D) Chemokine concentration profile for each tumor patch (row); Three 1-D heatmap (right) provides additional information for each patch: infiltration status (light/dark = infiltrated/deserted tumor), tumor cell (light/dark = tumor cells/ no tumor cells), and CD8+ T cells (light/dark = T cells/no T cells). 
  (E) Correlation between the chemokine signals and the presence of CD8+ T cells in a patch, yellow bars corresponds to chemokines that were discovered as key perturbations in panel A. 
  (F) UMAP projection of patches from melanoma tumor images (top), using only chemokine channels; colored arrows connect UMAP projection of patches without T cells and their corresponding perturbed image (bottom), color corresponds to k-nearest neighbor clusters of the perturbed images. Pie charts shows the patient distribution of original images near the four different clusters of perturbed images. 
  (G) Cell maps computed from one patient's IMC image, showing the predicted distribution of T cells before and after perturbation. 
  (H) Original vs. perturbed (predicted) median infiltration level across all patients (test group), error bar represents the interquartile range across multiple tumor sections per patient. Stage IV patients received perturbation strategy 1 (orange), stage III patients received perturbation strategy 2 (green). 
  (I) Predicted (median) infiltration level across all patients (test group) for different perturbation strategy of varying sparsity, error bar represents interquartile range.}
\end{figure}

Applying our counterfactual optimization procedure using the U-Net classifier trained on melanoma IMC images, we discovered a combinatorial therapy predicted to be highly effective in improving T cell infiltration in melanoma patients. For simplicity, we restricted the optimization algorithm to only perturb the level of chemokines, which is a family of secreted proteins that are known for their ability to stimulate cell migration \cite{hughes2018guide} and has been shown to play a significant role in T cell tumor localization \cite{steele2023t}. \autoref{fig:melanoma}A shows perturbations computed for each patient in the training group separates into two clusters. In patient cluster 1, only the level of CXCL9 is consistently increased, whereas in patient cluster 2, the levels of CXCL10 is consistently increased while the levels of CCL18 and CCL22 are consistently reduced (\autoref{fig:melanoma}A). Both CXCL9 and CXCL10 are well-known for playing a role in recruiting CD8+ T cells to tumor, whereas CCL22 is known to be a key chemokine for recruiting regulatory T cells, and relatively little is currently known regarding the role of CCL18 in T cell infiltration.

\autoref{fig:melanoma}B shows that the choice of which of the two strategies was selected for a patient appear to be strongly associated with the cancer stage of the patient, with strategy 1 being significantly enriched for stage IV colorectal cancer patients and strategy 2 being significantly enriched for stage III CRC patients. Probing deeper into the difference between these two patient clusters, we find that all chemokine are more highly expressed in the tumors of patients in cluster 2 compared to cluster 1, while there are no significant differences between the two groups in terms of the cell type compositions within tumors (\autoref{fig:melanoma}C). The fact that only patient cluster 2's strategy involves increasing CXCL10, but they actually have higher CXCL10 to start with suggests that differences in strategy between the two groups are not a simple reflection of differences in their chemokine levels, hinting at the possibility that there isn't a single optimal chemokine composition that supports T cell infiltration in melanoma tumor.

Our optimization procedure selected perturbations that would make the chemokine composition of a TME more similar to T cell rich regions of immune-infiltrated tumors. \autoref{fig:melanoma}D shows that melanoma tissue patches can be clustered into distinct groups based on their chemokine concentration profile. One cluster (highlighted by green box) contains exactly the patches from immune-infiltrated tumors that contains both tumor and T cells, which likely represents a chemokine signature that is suitable for T cell infiltration. On the other hand, a second cluster (highlighted by red box) which contains patches from immune-deserted tumors and has tumor but no T cell likely represents an unfavorable chemokine signature. In comparison to the second cluster highlighted in green, \autoref{fig:melanoma}B shows the first cluster (in red) from immune-infiltrated tumor contains elevated levels of CXCL9, CXCL10 and reduced levels of CCL22 which partially agrees with the perturbation strategy (\autoref{fig:melanoma}A) discovered by our optimization procedure. Notably, \autoref{fig:melanoma}E shows that our four selected chemokine targets cannot be predicted from simple correlation with the presence of CD8+ T cells.

We can directly observed how our optimization procedure searches for efficient perturbations by viewing both the original patch and perturbed patches in a dimensionality-reduced space. \autoref{fig:melanoma}F (top) shows in a UMAP space where each point represents the chemokine profile of an IMC patch, patches with CD8+ T cells are well-separated from those without CD8+ T cells. The colored arrows in the bottom panel of \autoref{fig:melanoma}F illustrating the perturbation for each patch as computed by our optimization procedure demonstrates two key features of our optimization procedure. First, the perturbations push patches without T cell towards the region in UMAP space as occupied by T-cell-infiltrated patches. Second, the colored arrows in \autoref{fig:melanoma}C show that the perturbations are efficient in that patches are perturbed just far enough to land in the desired region of space. Specifically, red points that start out on the right edge end up closer to the right after perturbation, while points that start on the left/bottom edge end up closer to the left/bottom, respectively. The colored patches are meant to illustrated this point. Furthermore, the pie charts (i-iv) show the region of space where points are being perturbed to are not occupied just tissue samples from a single patient with infiltrated tumor, rather these region consists of tissue samples from multiple patients, suggesting that our optimization procedure combines information from different patients when searching for perturbation strategies.

After applying in silico the set of five perturbations from \autoref{fig:melanoma}A to IMC image of a tumor, \autoref{fig:melanoma}G shows that T cells infiltration level (defined as \% of tumor cells with T cells nearby) is predicted to increase by 20 fold. Furthermore, \autoref{fig:melanoma}H shows that this predicted improvement in T cell infiltration holds across nearly all 14 patients from the test group, with little variability across multiple tumor samples of each patient. Specifically we selected which perturbation to perform \textit{in silico} based on the cancer stage of the patient. 

The combinatorial nature of our designed perturbation strategy is crucial to its predicted effectiveness. We systematically explore the importance of combinatorial perturbation by changing parameters $\beta$ of \autoref{eq:ldist} which adjusts the sparsity of the solution, where sparser solution means less molecules are perturbed. \autoref{fig:melanoma}I shows that perturbing multiple targets is predicted to be necessary for driving significant T cell infiltration across multiple patients. In conclusion, within the scope of the chemokine targets considered, combinatorial perturbation of the TME appears necessary for improving T cell infiltration.

\subsection*{Optimization framework predicts inhibition of CXCR4, CYR61, PD-1, and PD-L1 can improve T cell infiltration into liver tumor}

\begin{figure}[H]
  \centering
  \includegraphics[width=\textwidth]{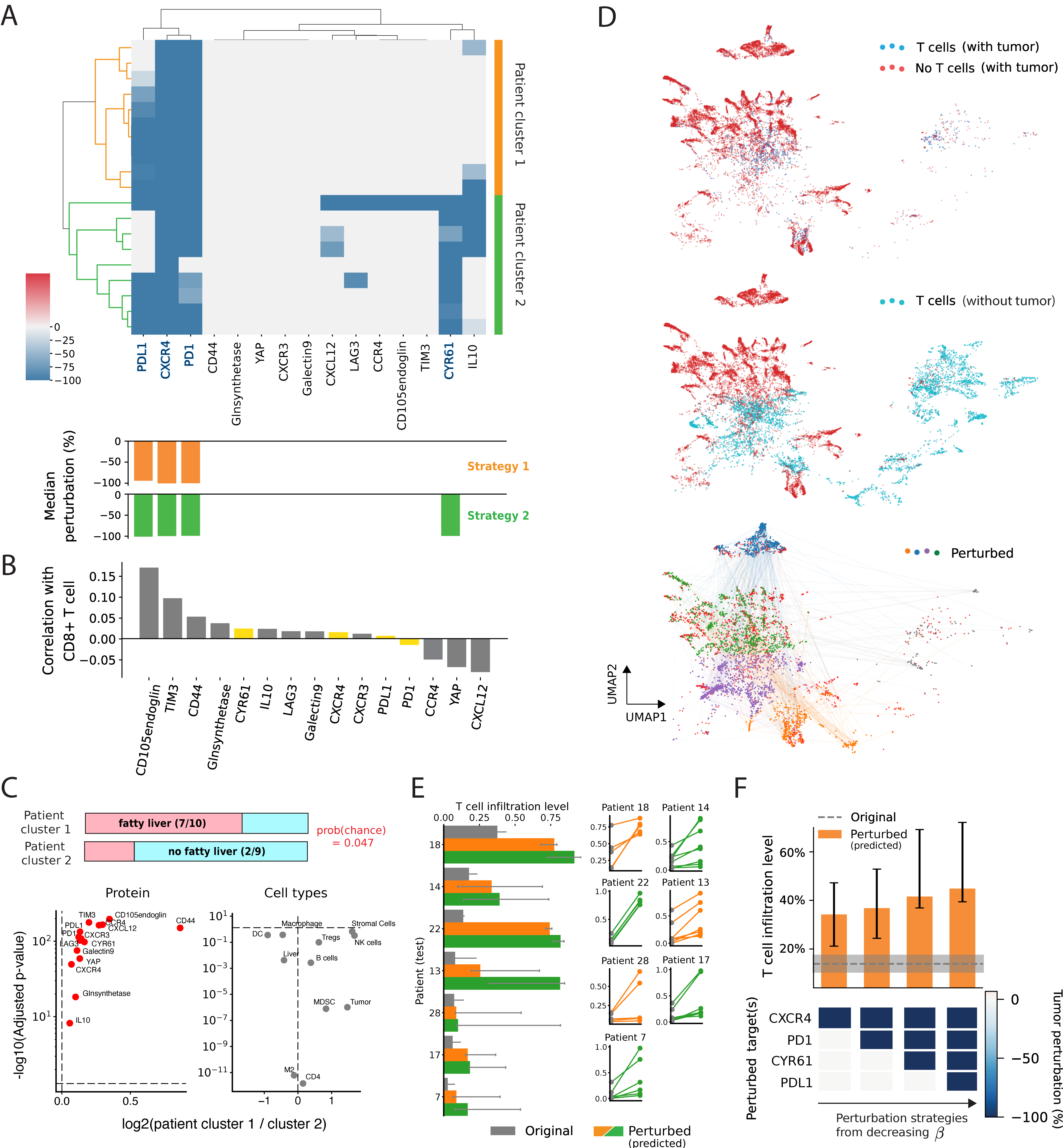}
  \caption{\textbf{Reducing levels of CXCR4, PD-1, CYR61 and IL10 predicted to drive T cell infiltration into immune-cold liver tumors.} 
  (A) Normalized whole-tumor perturbations optimized across IMC images of individual patients (row) from the train group. Green and orange marks to two patient clusters. 
  (B) Correlation between the protein level and the presence of CD8+ T cells. 
  (C) Horizontal bars indicate proportion of patients in each of the two groups that have fatty liver, where the probability of seeing such proportions by chance is $10\%$. T-tests with Bonferroni correction performed to assess differences in cell type composition and protein levels between the two patient cluster in (A), computed using median values and Wilcoxon rank sum test. 
  (D) UMAP projection of patches from liver tumor sections (top and middle); colored arrows connect UMAP projection of image patches without T cells and their corresponding optimally perturbed image (bottom), color corresponds to KNN clusters of the perturbed images.
  (E) Predict median T cell infiltration level in test population (left). Error bar represents the interquartile range across multiple tumor sections per patient. T cell infiltration level for individual tissue section before and after perturbation (right), where patient 13, 18, 28 have fatty liver disease. 
  (F) Predicted infiltration level across all test patients for different perturbation strategy of varying sparsity, error bar represents interquartile range.}
  \label{fig:liver}
\end{figure}

Applying our optimization framework to IMC images of CRC liver metastases, we discovered that a combination of 3-5 molecular perturbations is predicted to be highly effective in improving T cell infiltration. \autoref{fig:liver}A shows the optimal perturbations computed for every CRC patient in the training group, using IMC images of their liver metastases. Hierarchical clustering on this collection of perturbation strategies reveals our optimization framework consistently discovered two strategies for improving T cell infiltration from this patient population. The first cluster of patients, marked in green in \autoref{fig:liver}A, corresponds to a strategy that involves inhibiting PD-1, PD-L1, and CXCR4. The second strategy is similar except with the additional inhibition of CYR61. Interestingly, \autoref{fig:liver}B shows that nearly all of the perturbation targets selected correlated poorly with the presence of CD8+ T cells compared to the other proteins that were not selected as targets, suggesting the presence of significant spatial and nonlinear effects not captured by correlations alone. 

All targets identified by our method have been found to play crucial roles in suppressing effector T cell function in the TME, and inhibitors of certain targets have already been shown to drive T cell infiltration in CRC liver metastases. Regulatory T cells (Tregs) get recruited into tumor through CXCL12/CXCR4 interaction \cite{ghanem2014insights}. Next, the PD-1/PD-L1 pathway inhibits CD8+ T cell activity and infiltration in tumors. In addition, CYR61 is a chemoattractant and was recently shown to drive M2 TAM infiltration in patients with CRC liver metastases \cite{wang2023extracellular}. Inhibition of both PD-1 and CXCR4, which were consistently selected by our method as targets, have already been shown to increase CD8+ T cell infiltration in both CRC patients and mouse models \cite{biasci2020cxcr4, chen2015cxcr4, steele2023t}. Finally, \autoref{fig:liver}A shows the optimized therapy for a few patients involve inhibiting IL-10. As evidence in support of this prediction, blockade of IL-10 was recently shown to increase the frequency of non-exhausted CD8+ T cell in human colorectal cancer liver metastases \cite{sullivan2023blockade}. 

The fact that our method discovered two distinct perturbation strategies is partly a result of variation in liver fat build-up among patients. The first cluster is significantly enriched for patients with fatty liver disease ($70\%$ $(7/10)$) whereas only $22\%$ $(2/9)$ of patients in the second cluster have fatty liver disease ($p = 0.047$). Furthermore, \autoref{fig:liver}C shows that CYR61 level is significantly higher in the first group while having no significant difference in cell type composition. Indeed, CYR61 is known to be associated with non-alcoholic fatty liver disease (NAFLD). However despite patients in cluster 1 having higher levels of CYR61, it is only for patients in cluster 2 where the optimal strategy involves blocking CYR61. This result could be due to the fact that removing CYR61 from patients in cluster 1 is a much stronger perturbation due to their originally higher concentration. Such a large perturbation will likely push the tumor profile far out of distribution of the training data, making it an overall undesirable strategy.

\autoref{fig:liver}D (top) shows a UMAP projection of tumor patches either with or without T cells, note that in this case the T cell-present region tend to cluster together but is very low in abundance, suggesting it may be difficult to find perturbations that mimic such an environment. On the other hand, the middle UMAP figures shows that there are many region without tumor cells but contains T cells. Our optimization procedure appears to prioritize perturbing cells (green) to the region of space with tumor since the original images we are trying to optimize contain tumor cells as well, causing the green points to be most abundant (bottom panel \autoref{fig:liver}D). However, it appears to be a difficult target to perturb all the original images to, since in somes cases the algorithm resorted to perturbing images to be more similar to tissue patches that contain T cells but do not have tumors, as demonstrated by the presence of the purple and orange points. This results shows that our algorithm is flexible in that it does not always try to make tumor patches without T cells look like tumor patches with T cells. When doing so is difficult to achieve it can find alternative strategies that are also predicted to be effective.

The two therapeutic strategies we discovered generalizes to patients in our held-out test population. Given that we have two therapeutic strategies, we simulate in silico both perturbation strategies across all 7 test patients. \autoref{fig:liver}E (left) shows that for both treatment, CD8+ T cell infiltration level is predicted to increase significantly for nearly all patients, with the exception of patient $28$. Upon closer inspection of the perturbation effect on each tissue section, \autoref{fig:liver}E (right) shows the perturbation is predicted to significantly increase T cell infiltration level for only two out of the five tissue sections from patient $28$. This variability in response between tissue sections from the same patient is evident in nearly all test patients, suggesting that intra-tumor heterogeneity could be a significant issue in the development of personalized cancer therapy. Furthermore, \autoref{fig:liver}F shows again that combinatorial perturbation is necessary to drive significant T cell infiltration across multiple patients.

\section*{Discussion}


Our optimization framework combines deep neural networks with optimization methods from explainable AI to directly predict therapeutic interventions. 

One of the major strength of our model is that it can readily scale to deal with large diverse sets of patients samples, which will be crucial as more spatial transcriptomics and proteomics datasets are quickly becoming available \cite{chen2022spatiotemporal}. Larger data sets could allow us to train more complex models such as vision transformers, capturing long range interactions in tissue profiles that are may be more predictive of T cell localization. Furthermore, a large set of diverse patient samples will more accurately capture the extent of tumor heterogeneity, enabling our optimization framework to discover therapeutic strategies for different sub-classes of patients. 

For future work, we would like to apply our framework to spatial transcriptomics/proteomics data with hundreds to thousands of molecular channels. Currently, the number of spatial transcriptomics profiles of human tumor tissues is extremely limited, with most public data sets containing single tissue sections from 1-5 patients which is far too small for our application. However, as commercial platforms for spatial transcriptomics start to come online, we will likely be seeing large scale spatial transcriptomics data sets in the near future.

In this work, we focused on identifying generalized therapies by pooling predictions across multiple patient samples, but we can also apply our framework to find personalized therapy for treating individual patients. In both the melanoma and CRC liver metastases data sets, we see variability in the optimized perturbation between patients (\autoref{fig:melanoma}A, \autoref{fig:liver}A). Furthermore, \autoref{fig:liver}E shows that a therapeutic strategy could have highly variable effect across different tissue samples from the same patient. To generate therapy for individual patients, we would first obtain a large set of tissue samples from a patient's tumor and apply our optimization procedure to a random subset of the samples (training set), and then test the resulting perturbation strategy on the remaining samples (testing set) to see how well the strategy is predicted to perform across the entire tumor. 

Incorporating our model in closed loop with experimental data collection is another promising direction for future work. Data can be collected from patients or animal models with perturbed/engineered signaling context, and this data can be easily fed back into the classifier model to refine the model's prediction. The perturbation could be based on what the model predicts to be effective interventions, as is the case with our counterfactual optimization framework. We can also study the tissue samples on which the model tends to make the most mistake and train the model specifically using samples from similar sources, such as the same patient or disease state. 

For future work, we also plan to incorporate cell-type specific perturbations into our framework, which can be done by directly restricting the optimization to only alter signals within specific cell types. Experimentally, we plan to test our therapeutic predictions in murine models of melanoma and CRC liver metastases.

\section*{Methods}
\label{sec:methods}

\subsection*{Data split}
For both the Melanoma and Breast tumor dataset, we followed the same data splitting scheme to divide patients into three different groups (training, validation, testing) while ensuring similar class balance across the groups, which in our case means that the proportion of image patches with and without T cells are roughly equal across the three groups. Specifically, each image within a dataset was divided into $\SI{64}{\micro\meter} \times \SI{64}{\micro\meter}$ patches and the number of patches with and without CD8+ T cells was computed for each image. Next, the patients are shuffled between the three groups until three criteria are met: 1) the number of patients across the three groups follow a 65/15/20 ratio, 2) the difference in class proportion across the three groups is less than 2\%, and 3) the training set contains at least 65\% of total image patches.

\subsubsection*{Melanoma data set}
Melanoma data set from Hoch et al. contains 159 images/cores taken from 69 patients. Table below contains summary statistics for the three groups that patients were split into.
\begin{table}[h]
  \centering
  \begin{tabular}{lccc}
    \toprule
    Group  & Number of patients & Number of patches & Proportion of positive patches\\
    \midrule
    Training & $102 $ & $23741 $ & $29.6\%$  \\
    Validation & $28 $ & $6045 $ & $30.3\%$ \\
    Testing & $29$ & $5950 $ & $30.4\%$ \\
    \bottomrule
  \end{tabular}
  \caption{Data split statistic for Melanoma IMC dataset}
  \label{tab:melanoma_split}
\end{table}

\subsubsection*{Breast tumor data set}
Breast tumor dataset from Danenberg et al. contains 693 images/cores taken from 693 patients. Table below contains summary statistics for the three groups that patients were split into.
\begin{table}[h]
  \centering
  \begin{tabular}{lccc}
    \toprule
    Group  & Number of patients & Number of patches & Proportion of positive patches\\
    \midrule
    Training & $485$ & $41104 $ & $23.7\%$  \\
    Validation & $113$ & $9015 $ & $23.4\%$ \\
    Testing & $151$ & $12987 $ & $23.8\%$ \\
    \bottomrule
  \end{tabular}
  \caption{Data split statistic for breast tumor IMC dataset}
  \label{tab:breast_split}
\end{table}

\subsubsection*{CRC liver metastases data set}
Breast tumor dataset from Danenberg et al. contains 693 images/cores taken from 693 patients. Table below contains summary statistics for the three groups that patients were split into.
\begin{table}[h]
  \centering
  \begin{tabular}{lccc}
    \toprule
    Group  & Number of patients & Number of patches & Proportion of positive patches\\
    \midrule
    Training & $485$ & $41104 $ & $23.7\%$  \\
    Validation & $113$ & $9015 $ & $23.4\%$ \\
    Testing & $151$ & $12987 $ & $23.8\%$ \\
    \bottomrule
  \end{tabular}
  \caption{Data split statistic for breast tumor IMC dataset}
  \label{tab:liver_split}
\end{table}

\subsection*{Classifier training}
In this work, we trained three classes of models to perform our T cell prediction task. All models presented in this paper were trained with early stopping based on the validation Matthews Correlation Coefficient (MCC) for 10-20 epochs. All models were trained on an NVIDIA GeForce RTX 3090 Ti GPU using PyTorch version 1.13.1 \cite{paszke2019pytorch}. More details about hyperparameters and implementations can be found in our Github repository.

\subsubsection*{T cell masking strategy}
The purpose of model training is for the model to learn molecular features of the CD8+ T cell's environment that is indicative of its presence, so it is important for us to remove features of the image that are predictive of CD8+ T cell presence but are not part of the cell's environment. We devised a non-trivial cell masking strategy in order to remove T cell-intrinsic features without introducing new features that are highly predictive of T cell presence but are not biologically relevant. A simple masking strategy of zeroing out all pixels belonging to CD8+ T cells will introduce contiguous regions of zeros to image patches with T cells, which is an artificial feature that is nonetheless highly predictive of T cell presence and hence will likely be the main feature learned by a model during training. To circumvent this issue, we first apply a cell ``pixelation" step to the original IMC image where we reduce each cell to a single pixel positioned at the cell's centroid. The value of this pixel is the sum of all pixels originally associated with the cell, representing the total signal from each channel within the cell. We then mask this ``pixelated" image by zeroing all pixels representing CD8+ T cells. Since there are usually at most two T cell pixels in an image patch, zeroing them in a $16\times 16$ pixel image where most of the pixels ($>90\%$) are already zeroes is not likely to introduce a significant signal that is predictive T cell presence. We show that our strategy is effective at masking T cells without introducing additional features through a series of image augmentation experiments.

\subsubsection*{Logistic regression models}
We trained a single-layer neural network on the average intensity values from each molecular channel to obtain a logistic regression classifier, predicting the probability of CD8+ T cell presence in the image patch. This model represents a linear model where only the average intensity of each molecule is used for prediction instead of their spatial distribution within a patch.

\subsubsection*{MLP models}
Similar to a logistic regression model, the Multilayer Perceptron (MLP) also uses averaged intensity as input features for prediction but is capable of learning nonlinear interactions between features. The MLP model consists of two hidden layers (30 and 10 nodes) with ReLU activation.

\subsubsection*{U-Net models}
To train networks that can make full use of the spatial information, we used a fully convolutional neural network with the U-Net architecture. The U-Net architecture consists of a contracting path and an expansive path, which gives it a U-shaped structure \cite{buda2019association}. The contracting path consists of four repeated blocks, each containing a convolutional layer followed by a Rectified Linear Unit (ReLU) activation and a max pooling layer. The expansive path mirrors the contracting path, where each block contains a transposed convolutional layer. Skip connections concatenates the up-sampled features with the corresponding feature maps from the contracting path to include local information. The output of the expansive path is then fed to a fully-connected layer with softmax activation to produce a predicted probability. The model was trained from scratch using image augmentation to prevent over-fitting, including random horizontal/vertical flips and rotations, in addition to standard channel-wise normalization. We train our U-net classifiers using stochastic gradient descent with momentum and a learning rate of $10^{-2}$ on mini-batches of masked image patches.

\begin{table}[h]
  \caption{Performance of different classifier models trained on melanoma, liver tumor, and breast tumor IMC images to predict the presence of T cells ($p = 0.5$)}
  \centering
  \begin{tabular}{llcccccc}
    \toprule
    Cancer type  & Model & Accuracy & Precision & Recall & F1 & AUC-ROC & MCC\\
    \midrule
    Melanoma & Linear  &0.84& $0.83$  & $0.37$ & $0.51$ & $0.67$ & $0.48$ \\
    		 & MLP  &0.85& $0.79$ &  $0.48$ & $0.59$ & $0.72$ & $0.53$ \\
     		 & U-Net & 0.86 &  0.71& 0.61 & 0.65 & 0.77 & 0.57 \\
    Breast 	 & Linear 	&0.82& 0.57  & 0.16 & 0.24 & 0.57 & 0.23 \\
    tumor    	 & MLP 	&0.86& 0.71 &  0.42 & 0.52 & 0.69 & 0.47 \\
     		 & U-Net   & 0.87  & 0.70 &  0.48 & 0.56 & 0.71 & 0.50 \\
    Liver 	 & Linear 	& 0.86 & 0.71 & 0.19 & 0.29 & 0.59 & 0.31 \\
    metastasis  & MLP &  0.88& 0.67  & 0.50 & 0.57 & 0.73 & 0.51 \\
     		 & U-Net& 0.90 & 0.68 &  0.65 & 0.66 & 0.80 & 0.60 \\
    \bottomrule
  \end{tabular}
  \label{tab:metric}
\end{table}

\begin{table}[h]
  \caption{Performance of different CNN and ViT models trained on IMC image patches of melanoma}
  \centering
  \begin{tabular}{l|lccccc}
    \toprule
    Model Name  & U-Net & ResNet-18 & EfficientNet-B0 & MedViT \\
    \midrule
    Number of parameters  & 12.8M	& 11.2M & 4.1M & 31.3M  \\
    Test accuracy &  0.86	& 0.86 & 0.851 & 0.853  \\
    \bottomrule
  \end{tabular}
  \label{tab:modelcomparison}
\end{table}

\subsection*{Counterfactual optimization}
\label{sec:perturbation}

Given an IMC patch $x^{(i)}$ without T cells, and a classifier $f$, our goal is to find a perturbation $\delta^{(i)}$ for the patch such that $f$ classifies the perturbed patch as having T cells. For CNN models, $\delta^{(i)} \in \mathbb{R}^{w \times l \times d}$ is a 3D tensor that describes changes made for every channel, at each pixel of the patch. For MLP models, $\delta^{(i)} \in \mathbb{R}^d$ is a vector where each element represents changes to the average intensity of a channel in patch $i$. This distinction between the two models highlights differences in the type of perturbations that each model can discover. The MLP model can find region-wide perturbation such as changing the overall level of an extracellular molecule, whereas the CNN model allows for cell-type specific perturbations since perturbations are now being computed at the level of micron-scale pixels.

Given a CNN classifier $f$ and a IMC patch $x^{(i)}$ such that $f(x_0^{(i)}) = \mathbb{P}\text{(T cells present)} < p$, where $p > 0$ is the classification threshold below which the classifier predicts no T-cell, we aim to obtain a perturbation $\delta^{(i)}$ such that $f(x_0^{(i)} + \delta^{(i)}) > p$, by solving the following optimization problem adopted from \citep{looveren2021interpretable},

\begin{equation}
	\delta^{(i)} = \min_{\delta} L_\mathrm{pred}(x_0^{(i)}, \delta) + L_\mathrm{dist}(\delta) + L_\mathrm{proto}(x_0^{(i)}, \delta), \label{eq:objectivemethod} \\
\end{equation}
	such that 
\begin{align}
  L_\mathrm{pred}(x_0^{(i)}, \delta) &= c\max(-f(x_0^{(i)} + \delta),\, -p), \label{eq:lpred} \\
  L_\mathrm{dist}(\delta) &= \beta \lVert \delta \rVert_{1} + \lVert \delta \rVert_{2}^2, \label{eq:ldist} \\
  L_\mathrm{proto}(x_0^{(i)}, \delta) &= \theta  \lVert x_0^{(i)} + \delta -\mathrm{proto} \rVert_{2}^2, \label{proto} \\
  \delta^{(i)} &= \gamma^{(i)} \odot_3 x^{(i)}_0
\end{align}

where $\mathrm{proto}$ is an instance of the training set classified as having T cells, defined by first building a k-d tree of training instances classified as having T cells and setting the $k$-nearest item in the tree (in terms of euclidean distance to $x_0^{(i)}$) as $\mathrm{proto}$. We use $k=1$ for all optimization. 



\subsection*{Code Availability}
Code for model training, perturbation optimization and analysis are publicly available at \url{https://github.com/neonine2/deepspace}.

\subsection*{Data Availability}
This study used only published data sets that are publicly available.

\bibliography{bibliography}

\begin{thebibliography}{10}
\expandafter\ifx\csname url\endcsname\relax
  \def\url#1{\burl{#1}}\fi
\expandafter\ifx\csname urlprefix\endcsname\relax\def\urlprefix{URL }\fi
\providecommand{\bibinfo}[2]{#2}
\providecommand{\eprint}[2][]{\url{#2}}
\providecommand{\doi}[1]{\url{https://doi.org/#1}}
\bibcommenthead

\bibitem{fridman2017immune}
\bibinfo{author}{Fridman, W.~H.}, \bibinfo{author}{Zitvogel, L.}, \bibinfo{author}{Saut{\`e}s-Fridman, C.} \& \bibinfo{author}{Kroemer, G.}
\newblock \bibinfo{title}{The immune contexture in cancer prognosis and treatment}.
\newblock \emph{\bibinfo{journal}{Nature reviews Clinical oncology}} \textbf{\bibinfo{volume}{14}}~(12), \bibinfo{pages}{717--734} (\bibinfo{year}{2017}) .

\bibitem{binnewies2018understanding}
\bibinfo{author}{Binnewies, M.} \emph{et~al.}
\newblock \bibinfo{title}{Understanding the tumor immune microenvironment (time) for effective therapy}.
\newblock \emph{\bibinfo{journal}{Nature medicine}} \textbf{\bibinfo{volume}{24}}~(5), \bibinfo{pages}{541--550} (\bibinfo{year}{2018}) .

\bibitem{bruni2020immune}
\bibinfo{author}{Bruni, D.}, \bibinfo{author}{Angell, H.~K.} \& \bibinfo{author}{Galon, J.}
\newblock \bibinfo{title}{The immune contexture and immunoscore in cancer prognosis and therapeutic efficacy}.
\newblock \emph{\bibinfo{journal}{Nature Reviews Cancer}} \textbf{\bibinfo{volume}{20}}~(11), \bibinfo{pages}{662--680} (\bibinfo{year}{2020}) .

\bibitem{hegde2020top}
\bibinfo{author}{Hegde, P.~S.} \& \bibinfo{author}{Chen, D.~S.}
\newblock \bibinfo{title}{Top 10 challenges in cancer immunotherapy}.
\newblock \emph{\bibinfo{journal}{Immunity}} \textbf{\bibinfo{volume}{52}}~(1), \bibinfo{pages}{17--35} (\bibinfo{year}{2020}) .

\bibitem{choe2020engineering}
\bibinfo{author}{Choe, J.~H.}, \bibinfo{author}{Williams, J.~Z.} \& \bibinfo{author}{Lim, W.~A.}
\newblock \bibinfo{title}{Engineering t cells to treat cancer: the convergence of immuno-oncology and synthetic biology}.
\newblock \emph{\bibinfo{journal}{Annual Review of Cancer Biology}} \textbf{\bibinfo{volume}{4}}, \bibinfo{pages}{121--139} (\bibinfo{year}{2020}) .

\bibitem{pitt2016targeting}
\bibinfo{author}{Pitt, J.} \emph{et~al.}
\newblock \bibinfo{title}{Targeting the tumor microenvironment: removing obstruction to anticancer immune responses and immunotherapy}.
\newblock \emph{\bibinfo{journal}{Annals of Oncology}} \textbf{\bibinfo{volume}{27}}~(8), \bibinfo{pages}{1482--1492} (\bibinfo{year}{2016}) .

\bibitem{haslam2019estimation}
\bibinfo{author}{Haslam, A.} \& \bibinfo{author}{Prasad, V.}
\newblock \bibinfo{title}{Estimation of the percentage of us patients with cancer who are eligible for and respond to checkpoint inhibitor immunotherapy drugs}.
\newblock \emph{\bibinfo{journal}{JAMA network open}} \textbf{\bibinfo{volume}{2}}~(5), \bibinfo{pages}{e192535--e192535} (\bibinfo{year}{2019}) .

\bibitem{lee2019multiomics}
\bibinfo{author}{Lee, J.~S.} \& \bibinfo{author}{Ruppin, E.}
\newblock \bibinfo{title}{Multiomics prediction of response rates to therapies to inhibit programmed cell death 1 and programmed cell death 1 ligand 1}.
\newblock \emph{\bibinfo{journal}{JAMA oncology}} \textbf{\bibinfo{volume}{5}}~(11), \bibinfo{pages}{1614--1618} (\bibinfo{year}{2019}) .

\bibitem{pittet2022clinical}
\bibinfo{author}{Pittet, M.~J.}, \bibinfo{author}{Michielin, O.} \& \bibinfo{author}{Migliorini, D.}
\newblock \bibinfo{title}{Clinical relevance of tumour-associated macrophages}.
\newblock \emph{\bibinfo{journal}{Nature reviews Clinical oncology}} \textbf{\bibinfo{volume}{19}}~(6), \bibinfo{pages}{402--421} (\bibinfo{year}{2022}) .

\bibitem{bonaventura2019cold}
\bibinfo{author}{Bonaventura, P.} \emph{et~al.}
\newblock \bibinfo{title}{Cold tumors: a therapeutic challenge for immunotherapy}.
\newblock \emph{\bibinfo{journal}{Frontiers in immunology}} \textbf{\bibinfo{volume}{10}}, \bibinfo{pages}{168} (\bibinfo{year}{2019}) .

\bibitem{savas2016clinical}
\bibinfo{author}{Savas, P.} \emph{et~al.}
\newblock \bibinfo{title}{Clinical relevance of host immunity in breast cancer: from tils to the clinic}.
\newblock \emph{\bibinfo{journal}{Nature reviews Clinical oncology}} \textbf{\bibinfo{volume}{13}}~(4), \bibinfo{pages}{228--241} (\bibinfo{year}{2016}) .

\bibitem{tsaur2021immunotherapy}
\bibinfo{author}{Tsaur, I.}, \bibinfo{author}{Brandt, M.~P.}, \bibinfo{author}{Juengel, E.}, \bibinfo{author}{Manceau, C.} \& \bibinfo{author}{Ploussard, G.}
\newblock \bibinfo{title}{Immunotherapy in prostate cancer: new horizon of hurdles and hopes}.
\newblock \emph{\bibinfo{journal}{World journal of urology}} \textbf{\bibinfo{volume}{39}}, \bibinfo{pages}{1387--1403} (\bibinfo{year}{2021}) .

\bibitem{moffitt2022emerging}
\bibinfo{author}{Moffitt, J.~R.}, \bibinfo{author}{Lundberg, E.} \& \bibinfo{author}{Heyn, H.}
\newblock \bibinfo{title}{The emerging landscape of spatial profiling technologies}.
\newblock \emph{\bibinfo{journal}{Nature Reviews Genetics}} \textbf{\bibinfo{volume}{23}}~(12), \bibinfo{pages}{741--759} (\bibinfo{year}{2022}) .

\bibitem{lanitis2017mechanisms}
\bibinfo{author}{Lanitis, E.}, \bibinfo{author}{Dangaj, D.}, \bibinfo{author}{Irving, M.} \& \bibinfo{author}{Coukos, G.}
\newblock \bibinfo{title}{Mechanisms regulating t-cell infiltration and activity in solid tumors}.
\newblock \emph{\bibinfo{journal}{Annals of Oncology}} \textbf{\bibinfo{volume}{28}}, \bibinfo{pages}{xii18--xii32} (\bibinfo{year}{2017}) .

\bibitem{rodriques2019slide}
\bibinfo{author}{Rodriques, S.~G.} \emph{et~al.}
\newblock \bibinfo{title}{Slide-seq: A scalable technology for measuring genome-wide expression at high spatial resolution}.
\newblock \emph{\bibinfo{journal}{Science}} \textbf{\bibinfo{volume}{363}}~(6434), \bibinfo{pages}{1463--1467} (\bibinfo{year}{2019}) .

\bibitem{eng2019transcriptome}
\bibinfo{author}{Eng, C.-H.~L.} \emph{et~al.}
\newblock \bibinfo{title}{Transcriptome-scale super-resolved imaging in tissues by rna seqfish+}.
\newblock \emph{\bibinfo{journal}{Nature}} \textbf{\bibinfo{volume}{568}}~(7751), \bibinfo{pages}{235--239} (\bibinfo{year}{2019}) .

\bibitem{giesen2014highly}
\bibinfo{author}{Giesen, C.} \emph{et~al.}
\newblock \bibinfo{title}{Highly multiplexed imaging of tumor tissues with subcellular resolution by mass cytometry}.
\newblock \emph{\bibinfo{journal}{Nature methods}} \textbf{\bibinfo{volume}{11}}~(4), \bibinfo{pages}{417--422} (\bibinfo{year}{2014}) .

\bibitem{goltsev2018deep}
\bibinfo{author}{Goltsev, Y.} \emph{et~al.}
\newblock \bibinfo{title}{Deep profiling of mouse splenic architecture with codex multiplexed imaging}.
\newblock \emph{\bibinfo{journal}{Cell}} \textbf{\bibinfo{volume}{174}}~(4), \bibinfo{pages}{968--981} (\bibinfo{year}{2018}) .

\bibitem{bhate2022tissue}
\bibinfo{author}{Bhate, S.~S.}, \bibinfo{author}{Barlow, G.~L.}, \bibinfo{author}{Sch{\"u}rch, C.~M.} \& \bibinfo{author}{Nolan, G.~P.}
\newblock \bibinfo{title}{Tissue schematics map the specialization of immune tissue motifs and their appropriation by tumors}.
\newblock \emph{\bibinfo{journal}{Cell Systems}} \textbf{\bibinfo{volume}{13}}~(2), \bibinfo{pages}{109--130} (\bibinfo{year}{2022}) .

\bibitem{wu2022graph}
\bibinfo{author}{Wu, Z.} \emph{et~al.}
\newblock \bibinfo{title}{Graph deep learning for the characterization of tumour microenvironments from spatial protein profiles in tissue specimens}.
\newblock \emph{\bibinfo{journal}{Nature Biomedical Engineering}} \bibinfo{pages}{1--14} (\bibinfo{year}{2022}) .

\bibitem{schurch2020coordinated}
\bibinfo{author}{Sch{\"u}rch, C.~M.} \emph{et~al.}
\newblock \bibinfo{title}{Coordinated cellular neighborhoods orchestrate antitumoral immunity at the colorectal cancer invasive front}.
\newblock \emph{\bibinfo{journal}{Cell}} \textbf{\bibinfo{volume}{182}}~(5), \bibinfo{pages}{1341--1359} (\bibinfo{year}{2020}) .

\bibitem{hoch2022multiplexed}
\bibinfo{author}{Hoch, T.} \emph{et~al.}
\newblock \bibinfo{title}{Multiplexed imaging mass cytometry of the chemokine milieus in melanoma characterizes features of the response to immunotherapy}.
\newblock \emph{\bibinfo{journal}{Science Immunology}} \textbf{\bibinfo{volume}{7}}~(70), \bibinfo{pages}{eabk1692} (\bibinfo{year}{2022}) .

\bibitem{wang2023extracellular}
\bibinfo{author}{Wang, Z.} \emph{et~al.}
\newblock \bibinfo{title}{Extracellular vesicles in fatty liver promote a metastatic tumor microenvironment}.
\newblock \emph{\bibinfo{journal}{Cell Metabolism}}  (\bibinfo{year}{2023}) .

\bibitem{danenberg2022breast}
\bibinfo{author}{Danenberg, E.} \emph{et~al.}
\newblock \bibinfo{title}{Breast tumor microenvironment structures are associated with genomic features and clinical outcome}.
\newblock \emph{\bibinfo{journal}{Nature genetics}} \textbf{\bibinfo{volume}{54}}~(5), \bibinfo{pages}{660--669} (\bibinfo{year}{2022}) .

\bibitem{buda2019association}
\bibinfo{author}{Buda, M.}, \bibinfo{author}{Saha, A.} \& \bibinfo{author}{Mazurowski, M.~A.}
\newblock \bibinfo{title}{{Association of genomic subtypes of lower-grade gliomas with shape features automatically extracted by a deep learning algorithm}}.
\newblock \emph{\bibinfo{journal}{Computers in Biology and Medicine}} \textbf{\bibinfo{volume}{109}} (\bibinfo{year}{2019}).
\newblock \doi{10.1016/j.compbiomed.2019.05.002} .

\bibitem{looveren2021interpretable}
\bibinfo{author}{Looveren, A.~V.} \& \bibinfo{author}{Klaise, J.}
\newblock \bibinfo{title}{Interpretable counterfactual explanations guided by prototypes}.
\newblock \emph{\bibinfo{journal}{Joint European Conference on Machine Learning and Knowledge Discovery in Databases}} \bibinfo{pages}{650--665} (\bibinfo{year}{2021}) .

\bibitem{hughes2018guide}
\bibinfo{author}{Hughes, C.~E.} \& \bibinfo{author}{Nibbs, R.~J.}
\newblock \bibinfo{title}{A guide to chemokines and their receptors}.
\newblock \emph{\bibinfo{journal}{The FEBS journal}} \textbf{\bibinfo{volume}{285}}~(16), \bibinfo{pages}{2944--2971} (\bibinfo{year}{2018}) .

\bibitem{steele2023t}
\bibinfo{author}{Steele, M.~M.} \emph{et~al.}
\newblock \bibinfo{title}{T cell egress via lymphatic vessels is tuned by antigen encounter and limits tumor control}.
\newblock \emph{\bibinfo{journal}{Nature Immunology}} \textbf{\bibinfo{volume}{24}}~(4), \bibinfo{pages}{664--675} (\bibinfo{year}{2023}) .

\bibitem{ghanem2014insights}
\bibinfo{author}{Ghanem, I.} \emph{et~al.}
\newblock \bibinfo{title}{{Insights on the CXCL12-CXCR4 axis in hepatocellular carcinoma carcinogenesis}}.
\newblock \emph{\bibinfo{journal}{{American journal of translational research}}} \textbf{\bibinfo{volume}{6}}~(4), \bibinfo{pages}{340} (\bibinfo{year}{2014}) .

\bibitem{biasci2020cxcr4}
\bibinfo{author}{Biasci, D.} \emph{et~al.}
\newblock \bibinfo{title}{{CXCR4 inhibition in human pancreatic and colorectal cancers induces an integrated immune response}}.
\newblock \emph{\bibinfo{journal}{{Proceedings of the National Academy of Sciences}}} \textbf{\bibinfo{volume}{117}}~(46), \bibinfo{pages}{28960--28970} (\bibinfo{year}{2020}) .

\bibitem{chen2015cxcr4}
\bibinfo{author}{Chen, Y.} \emph{et~al.}
\newblock \bibinfo{title}{Cxcr4 inhibition in tumor microenvironment facilitates anti-programmed death receptor-1 immunotherapy in sorafenib-treated hepatocellular carcinoma in mice}.
\newblock \emph{\bibinfo{journal}{Hepatology}} \textbf{\bibinfo{volume}{61}}~(5), \bibinfo{pages}{1591--1602} (\bibinfo{year}{2015}) .

\bibitem{sullivan2023blockade}
\bibinfo{author}{Sullivan, K.~M.} \emph{et~al.}
\newblock \bibinfo{title}{{Blockade of interleukin 10 potentiates antitumour immune function in human colorectal cancer liver metastases}}.
\newblock \emph{\bibinfo{journal}{Gut}} \textbf{\bibinfo{volume}{72}}~(2), \bibinfo{pages}{325--337} (\bibinfo{year}{2023}) .

\bibitem{chen2022spatiotemporal}
\bibinfo{author}{Chen, A.} \emph{et~al.}
\newblock \bibinfo{title}{Spatiotemporal transcriptomic atlas of mouse organogenesis using dna nanoball-patterned arrays}.
\newblock \emph{\bibinfo{journal}{Cell}} \textbf{\bibinfo{volume}{185}}~(10), \bibinfo{pages}{1777--1792} (\bibinfo{year}{2022}) .

\bibitem{paszke2019pytorch}
\bibinfo{author}{Paszke, A.} \emph{et~al.}
\newblock \bibinfo{title}{{PyTorch: An imperative style, high-performance deep learning library}}.
\newblock \emph{\bibinfo{journal}{{Advances in neural information processing systems}}} \textbf{\bibinfo{volume}{32}} (\bibinfo{year}{2019}) .

\end{thebibliography}

\subsection*{Acknowledgements}
We would like to thank Inna Strazhnik for her support with figure illustrations. We would like to thank Akil Merchant, Alma Andersson, Aviv Regev, Long Cai, Barbara Wold, Michal Polonsky, Jonathan Fox, Yujing Yang, Abdullah Farooq and all members of the Thomson lab for insightful discussion that significantly improved this work.

\end{document}